\documentclass{jetpl}

\usepackage{graphicx}

\twocolumn \lat
\begin{document}

\title{Manifestation of geometric effects in temperature behavior of AC magnetic response
of Josephson Junction Arrays} \rtitle{} \sodtitle{Manifestation of
geometric effects in Josephson Junction Arrays}

\author{S. Sergeenkov$^{1,2}$ and F.M. Araujo-Moreira$^{1}$}

\address{$^{1}$ Departamento de F\'{i}sica e Engenharia F\'{i}sica,
Grupo de Materiais e Dispositivos,\\ Centro Multidisciplinar para
o Desenvolvimento de Materiais Cer\^amicos,\\ Universidade Federal
de S\~ao Carlos,
S\~ao Carlos, SP, 13565-905 Brazil\\
$^{2}$ Bogoliubov Laboratory of Theoretical Physics, Joint
Institute for Nuclear Research,\\ Dubna 141980, Moscow Region,
Russia}



\abstract{ By improving resolution of home-made mutual-inductance
measurements technique, a pronounced step-like structure (with the
number of steps $n=4$ for all AC fields) has been observed in the
temperature dependence of AC susceptibility in artificially
prepared two-dimensional Josephson Junction Arrays (2D-JJA) of
unshunted $Nb-AlO_x-Nb$ junctions with $\beta _L(4.2K)=30$. Using
a single-plaquette approximation of the overdamped 2D-JJA model,
we were able to successfully fit our data assuming that steps are
related to the geometric properties of the plaquette. The number
of steps $n$ corresponds to the number of flux quanta that can be
screened by the maximum critical current of the junctions. The
steps are predicted to manifest themselves in arrays with the
inductance related parameter $\beta _L(T)$ matching a
"quantization" condition $\beta _L(0)=2\pi (n+1)$.}

\PACS{74.25.Ha, 74.50.+r, 74.80.Bj}

\maketitle

{\bf 1. Introduction.} Many unusual and still not completely
understood magnetic properties of Josephson Junction Arrays (JJAs)
continue to attract attention of both theoreticians and
experimentalists alike (for recent reviews on the subject see,
e.g.~\cite{1,2,3,4} and further references therein). In
particular, among the numerous spectacular phenomena recently
discussed and observed in JJAs we would like to mention the
dynamic temperature reentrance of AC susceptibility~\cite{2}
(closely related to paramagnetic Meissner effect~\cite{3}) and
avalanche-like magnetic field behavior of magnetization~\cite{4,5}
(closely related to self-organized criticality (SOC)~\cite{6,7}).
More specifically, using highly sensitive SQUID magnetometer,
magnetic field jumps in the magnetization curves associated with
the entry and exit of avalanches of tens and hundreds of fluxons
were clearly seen in SIS-type arrays~\cite{5}. Besides, it was
shown that the probability distribution of these processes is in
good agreement with the SOC theory~\cite{7}. An avalanche
character of flux motion was observed at temperatures at which the
size of the fluxons did not exceed the size of the cell, that is,
for discrete vortices. On the other hand, using a similar
technique, magnetic flux avalanches were not observed in SNS-type
proximity arrays~\cite{8} despite a sufficiently high value of the
inductance $L$ related critical parameter $\beta _L=2\pi LI_C/\Phi
_0$ needed to satisfy the observability conditions of SOC.
Instead, the observed quasi-hydrodynamic flux motion in the array
was explained by the considerable viscosity characterizing the
vortex motion through the Josephson junctions.

In this paper we present experimental evidence for manifestation
of novel geometric effects in magnetic response of high-quality
ordered 2D-JJA. By increasing the resolution of our home-made
mutual-inductance measurements technique, we were able to observe
for the first time a fine, step-like structure (with the number of
steps $n=4$ for all AC fields used in our experiments) in the
temperature behavior of AC susceptibility in artificially prepared
2D-JJA of unshunted $Nb-AlO_x-Nb$ junctions. Using a
single-plaquette approximation of the overdamped 2D-JJA model, we
show that the number of steps $n$ corresponds to the number of
flux quanta that can be screened by the maximum critical current
of the junctions and as a result steps will manifest themselves in
arrays with the inductance related parameter $\beta _L(T)$
matching a "quantization" condition $\beta _L(0)=2\pi (n+1)$.

{\bf 2. Experimental results.} To measure the complex AC
susceptibility in our arrays with high precision,  we used a
home-made susceptometer based on the so-called screening method in
the reflection configuration~\cite{9,10,11}. The experimental
system was calibrated by using a high-quality niobium thin film.
Previously, we have shown that the calibrated output of the
complex voltage in this experimental setup corresponds to the true
complex AC susceptibility (for more details on the experimental
technique and setups used in our study, see~\cite{2,11}).

Measurements were performed as a function of the temperature $T$ (for $%
1.5K<T<15K$), and the amplitude of the excitation field $h_{ac}$ (for $%
1mOe<h_{ac}<10Oe$) normal to the plane of the array. The frequency
of AC field in the experiments reported here was fixed at $20kHz$.
The used in the present study unshunted 2D-JJAs are formed by
loops of niobium islands (with $T_C=9.25K$) linked through
$Nb-AlO_{x}-Nb$ Josephson junctions and consist of $100\times 150$
tunnel junctions. The unit cell has square geometry with lattice
spacing $a=46\mu m$ and a single junction area of $5\times 5\mu
m^{2}$. Since the inductance of each loop is $L=\mu _{0}a = 64pH$
and the critical current of each junction is $I_{C}(4.2K)= 150\mu
A$, we have $\beta _{L}(4.2K)=30$. Recall that parameter $\beta
_{L}(T)=2\pi LI_{C}(T)/\Phi _{0}$ (where $\Phi _0$ is the magnetic
flux quantum) is proportional to the number of flux quanta that
can be screened by the maximum critical current in the junctions.

It is important to mention that magnetic field dependence of the
critical current of the array (taken at $T=4.2K$) on DC magnetic
field $H_{dc}$ (parallel to the plane of the sample)
exhibited~\cite{2,11} a sharp Fraunhofer-like pattern
characteristic of a single-junction response, thus proving a
rather strong coherence within arrays (with negligible
distribution of critical currents and sizes of the individual
junctions) and hence the high quality of our samples.

\begin{figure}
\centerline{\includegraphics[width=80mm]{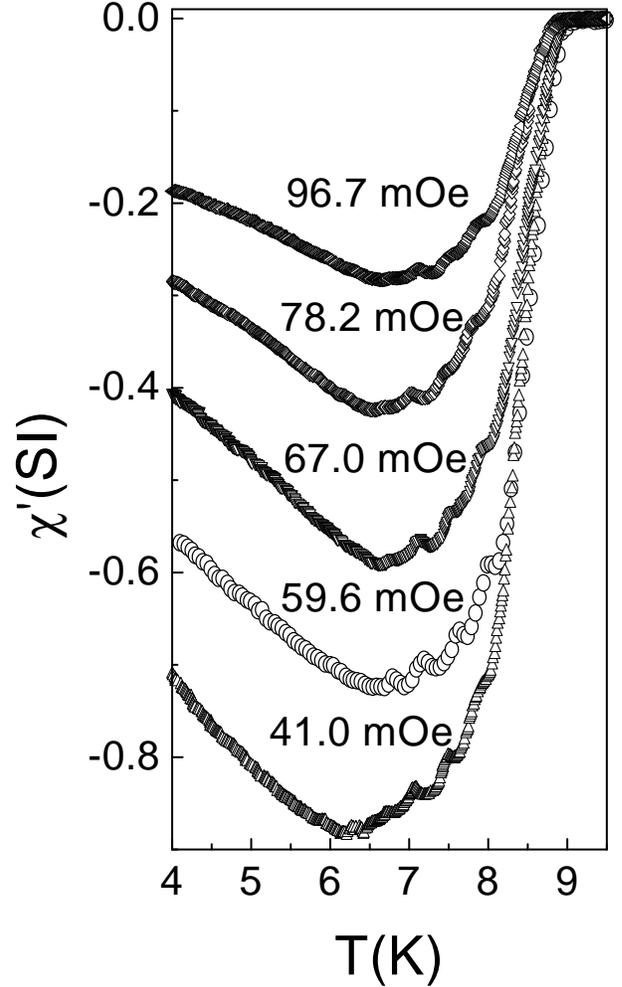}} \vspace{0.5cm}
 \caption{Fig.\ref{fig:fig1}. Experimental results for temperature dependence of the
real part of AC susceptibility $\chi '(T,h_{ac})$ for different AC
field amplitudes $h_{ac}=41.0$, $59.6$, $67.0$, $78.2$ and $96.7
mOe$. } \label{fig:fig1}
\end{figure}

The observed temperature dependence of the real part of AC
susceptibility for different AC fields is shown in
Fig.~\ref{fig:fig1}. A pronounced step-like structure is clearly
seen at higher temperatures. The number of steps $n$ does not
depend on AC field amplitude and is equal to $n=4$. As
expected~\cite{2,11,12}, for $h_{ac}>40mOe$ (when the array is in
the mixed-like state with practically homogeneous flux
distribution) the steps are accompanied by the previously observed
reentrant behavior with $\chi ^{\prime }(T,h_{ac})$ starting to
increase at low temperatures.

{\bf 3. Discussion.} To understand the step-like behavior of the
AC susceptibility observed in unshunted 2D-JJAs, in principle one
would need to analyze in detail the flux dynamics in these arrays.
However, as we have previously reported~\cite{2,11,12}, because of
the well-defined periodic structure of our arrays with no visible
distribution of junction sizes and critical currents, it is quite
reasonable to assume that the experimental results obtained from
the magnetic properties of our 2D-JJAs could be understood by
analyzing the dynamics of just a single unit cell (plaquette) of
the array. As we shall see, theoretical interpretation of the
presented here experimental results based on single-loop
approximation, is in excellent agreement with the observed
behavior. In our analytical calculations, the unit cell is a loop
containing four identical Josephson junctions and the measurements
correspond to the zero-field cooling AC magnetic susceptibility.
If we apply an AC external field $H_{ac}(t)=h_{ac}\cos \omega t$
normally to the 2D-JJA, then the total magnetic flux $\Phi (t)$
threading the four-junction superconducting loop is given by $\Phi
(t)=\Phi _{ext}(t)+LI(t)$ where $L$ is the loop inductance, $\Phi
_{ext}(t)=SH_{ac}(t)$ is the flux related to the applied magnetic
field (with $S\simeq a^2$ being the projected area of the loop),
and the circulating current in the loop reads $I(t)=I_C(T)\sin
\phi (t)$. Here $\phi (t)$ is the gauge-invariant superconducting
phase difference across the $i$th junction. As is well-known, in
the case of four junctions, the flux quantization condition
reads~\cite{11,13}
\begin{equation}
\phi =\frac{\pi }{2}\left( n+\frac{\Phi }{\Phi _{0}}\right )
\end{equation}
where $n$ is an integer and for simplicity we assume as usual
that~\cite{2,11} $\phi _{1}=\phi _{2}=\phi _{3}=\phi _{4}\equiv
\phi $.

To properly treat the magnetic properties of the system, let us
introduce the following Hamiltonian
\begin{equation}
{\cal H}(t)=J(T)[1-\cos \phi (t)]+\frac{1}{2}LI^2(t)
\end{equation}
which describes the tunneling (first term) and inductive (second
term) contributions to the total energy of a single plaquette.
Here, $J(T)=(\Phi _0/2\pi )I_C(T)$ is the Josephson coupling
energy.

\begin{figure}
\centerline{\includegraphics[width=90mm]{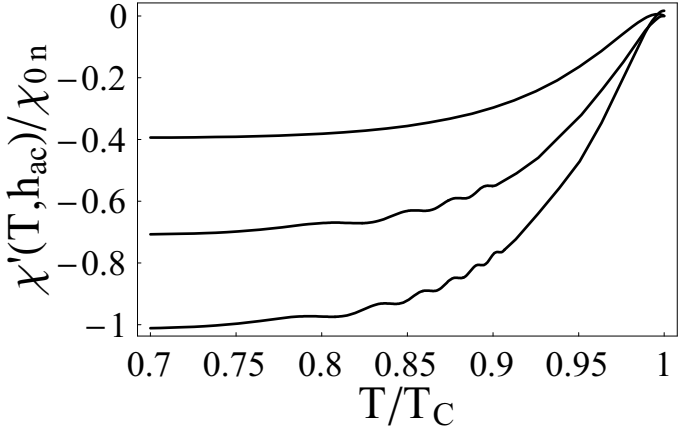}} \vspace{0.5cm}
 \caption{Fig.\ref{fig:fig2}. Theoretically predicted dependence
 of the normalized susceptibility $\chi ^{\prime}(T,h_{ac})/\chi _{0n}$
 (for better visual effect the curves are normalized
differently with  $\chi _{00}=2.5\chi _{0}$, $\chi _{03}=1.5\chi
_{0}$, and $\chi _{05}=\chi _{0}$ where $\chi _0=S^2/VL$) on
reduced temperature $T/T_C$ according to Eqs.(3)-(5) for $f=0.5$
and  for "quantized" values of $\beta _L(0)=2\pi (n+1)$ (from top
to bottom): $n=0$, $3$ and $5$.}
 \label{fig:fig2}
\end{figure}

Before turning to the interpretation of the observed step-like
structure of $\chi ^{\prime}(T,h_{ac})$, we would like to briefly
comment on the origin of reentrant behavior in our unshunted
arrays (which has been discussed in much detail earlier,
see~\cite{2,12}). A comparative study of the magnetic properties
of two-dimensional arrays of both unshunted and shunted (using a
molybdenum shunt resistor short-circuiting each junction)
$Nb-AlO_x-Nb$ junctions revealed~\cite{12} that the dynamic
reentrance phenomenon can be explained by properly addressing the
(neglected in our overdamped model) damping effects associated
with the finite value of the Stewart-McCumber parameter $\beta
_C(T)=2\pi C_JR_J^2I_C(T)/\Phi _0$ (where $C_{J}$ is the
capacitance and $R_{J}$ is the quasi-particle resistance of the
array). More precisely, the reentrance was found to take place in
considered here unshunted arrays (with $\beta _C(4.2K)=30$) and
totally disappeared in shunted arrays (with $\beta _C(4.2K)=1$).
It is important to mention that both arrays had the same value of
the $\beta _{L}$ parameter, namely $\beta _{L}(4.2K)=30$.

Returning to the discussion of the observed here geometrical
effects, we notice that the number of observed steps $n$ (in our
case $n=4$) clearly hints at a possible connection between the
observed here phenomenon and flux quantization condition within a
single four-junction plaquette. Indeed, the circulating in the
loop current $I(t)=I_C(T)\sin \phi (t)$ passes through its maximum
value whenever $\phi (t)$ reaches the value of
$\frac{\pi}{2}(2n+1)$ with $n=0,1,2..$. As a result, the maximum
number of fluxons threading a single plaquette (see Eq.(1)) over
the period $2\pi /\omega$ becomes equal to $<\Phi (t)>=(n+1)\Phi
_{0}$. In turn, the latter equation is equivalent to the following
condition $\beta _L(T)=2\pi (n+1)$. Since this formula is valid
for any temperature, we can rewrite it as a geometrical
"quantization" condition $\beta _L(0)=2\pi (n+1)$. Recall that in
our array $\beta _{L}(0)=31.6$ (extrapolated from its experimental
value $\beta _{L}(4.2K)=30$) which is a perfect match for the
above "quantization" condition predicting $n=4$ for the number of
steps in a single plaquette, in excellent agreement with the
observations.

Based on the above discussion, we conclude that in order to
reproduce the observed temperature steps in the behavior of AC
susceptibility, we need a particular solution to Eq.(1) for the
phase difference in the form of $\phi
_n(t)=\frac{\pi}{2}(2n+1)+\delta \phi (t)$ assuming $\delta \phi
(t)\ll 1$. After substituting this Ansatz into Eq.(1), we find
that $\phi _n(t)\simeq \frac{\pi}{2}n+\frac{1}{4}\beta
_L(T)+\frac{1}{4}f\cos \omega t$ where $f=2\pi Sh_{ac}/\Phi _0$ is
the AC field related frustration parameter. Using this effective
phase difference, we can calculate the AC response of a single
plaquette. Namely, the real part of susceptibility reads
\begin{equation}
\chi '(T,h_{ac})=\frac{1}{\pi}\int_0^{\pi}d(\omega t)\cos (\omega
t) \chi _n(t)
\end{equation}
where
\begin{equation}
\chi _n(t)=-\frac{1}{V}\left [\frac{\partial ^2{\cal H}}{\partial
h^2_{ac}}\right ]_{\phi =\phi _n(t)}
\end{equation}
Here $V$ is the sample's volume.

For the explicit temperature dependence of $\beta _{L}(T)=2\pi
LI_{C}(T)/\Phi _{0}$ we use the well-known~\cite{14,15} analytical
approximation of the BCS gap parameter (valid for all
temperatures), $\Delta (T)=\Delta (0)\tanh
\left(2.2\sqrt{\frac{T_{C}-T}{T}}\right)$ with $\Delta
(0)=1.76k_BT_C$ which governs the temperature dependence of the
Josephson critical current
\begin{equation}
I_{C}(T)=I_{C}(0)\left[ \frac{\Delta (T)}{\Delta (0)}\right] \tanh
\left[ \frac{\Delta (T)}{2k_{B}T}\right]
\end{equation}
Fig.~\ref{fig:fig2} depicts the predicted by Eqs.(3)-(5)
dependence of the AC susceptibility on reduced temperature for
$f=0.5$ and for different "quantized" values of $\beta _L(0)=2\pi
(n+1)$. Notice the appearance of three and five steps for $n=3$
and $n=5$, respectively (as expected, case $n=0$ corresponds to a
smooth temperature behavior without steps).

\begin{figure}
\centerline{\includegraphics[width=80mm]{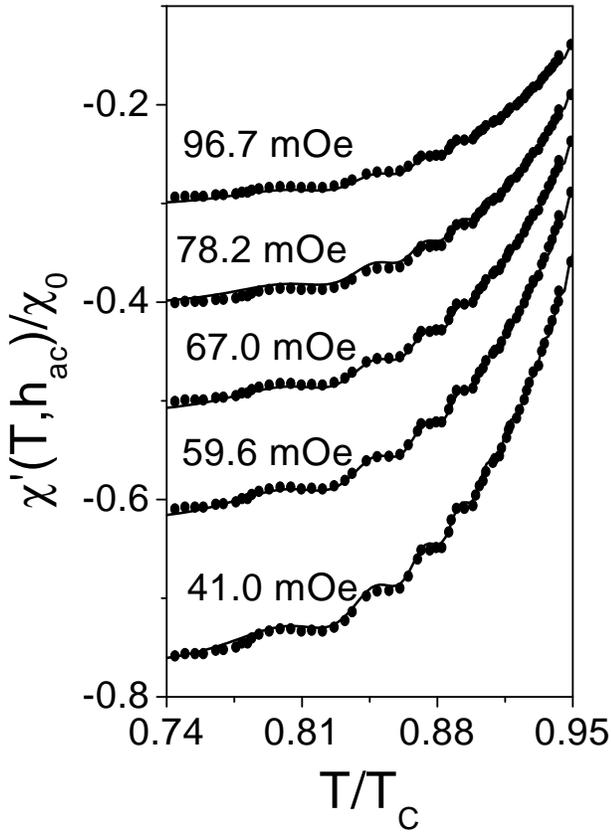}} \vspace{0.5cm}
 \caption{Fig.\ref{fig:fig3}. Fits (solid lines) of the experimental data for $h_{ac}
 =41.0$, $59.6$, $67.0$, $78.2$, and $96.7 mOe$ according to
Eqs.(3)-(5) with $\beta _{L}(0)=10\pi$.} \label{fig:fig3}
\end{figure}

In Fig.~\ref{fig:fig3} we present fits (shown by solid lines) of
the observed temperature dependence of the normalized
susceptibility $\chi ^{\prime}(T,h_{ac})/\chi _0$ for different
magnetic fields $h_{ac}$ according to Eqs.(3)-(5) using $\beta
_{L}(0)=10\pi$. As is seen, our simplified model based on a
single-plaquette approximation demonstrates an excellent agreement
with the observations.

In summary, a step-like structure (accompanied by previously seen
low-temperature reentrance phenomenon) has been observed for the
first time in the temperature dependence of AC susceptibility in
artificially prepared two-dimensional Josephson Junction Arrays of
unshunted $Nb-AlO_x-Nb$ junctions. The steps are shown to occur in
arrays with the inductance related parameter $\beta _L(T)$
matching the "quantization" condition $\beta _L(0)=2\pi (n+1)$
where $n$ is the number of steps.

We thank W. Maluf for his help in running the experiments. We are
indebted to P. Barbara, C.J. Lobb, R.S. Newrock, and A. Sanchez
for useful discussions. The authors gratefully acknowledge
financial support from Brazilian Agency FAPESP under grant
2003/00296-5.

\end{document}